\documentclass[prl,aps,twocolumn,showpacs,superscriptaddress]{revtex4}
\usepackage{graphicx}
\usepackage{bm}

\newcommand{\beq}{\begin{equation}}
\newcommand{\eeq}{\end{equation}}
\newcommand{\barr}{\begin{eqnarray}}
\newcommand{\earr}{\end{eqnarray}}

\begin{document}

\title{ Dynamical Imperfections in quantum computers}
\author{Paolo Facchi}
\affiliation{Dipartimento di Fisica, Universit\`a di Bari,
        I-70126  Bari, Italy \\
            and INFN, Sezione di Bari, I-70126 Bari, Italy}
\author{Simone Montangero}
\affiliation{NEST-INFM $\&$ Scuola Normale Superiore, Piazza
        dei Cavalieri 7, 56126 Pisa, Italy}
\homepage{http://www.sns.it/QTI/}
\author{Rosario Fazio}
\affiliation{NEST-INFM $\&$ Scuola Normale Superiore, Piazza
        dei Cavalieri 7, 56126 Pisa, Italy}
\homepage{http://www.sns.it/QTI/}
\author{Saverio Pascazio}
\affiliation{Dipartimento di Fisica, Universit\`a di Bari,
        I-70126  Bari, Italy \\
            and INFN, Sezione di Bari, I-70126 Bari, Italy}

\begin{abstract}
We study the effects of dynamical imperfections in quantum
computers. By considering an explicit example, we identify different
regimes ranging from the low-frequency case, where the imperfections
can be considered as static but with renormalized parameters, to the
high frequency fluctuations, where the effects of imperfections are
completely wiped out. We generalize our results by proving a theorem
on the dynamical evolution of a system in the presence of dynamical
perturbations.
\end{abstract}

\date{\today}
\pacs{03.67.Lx, 05.45.Mt, 24.10.Cn, 03.67.Mn, 03.67.}

\maketitle

In any experimental implementation of a quantum information
protocol~\cite{nielsen} one  has to face the presence of errors. The
coupling of the quantum computer to the surrounding environment is
responsible for decoherence~\cite{chuang95} which ultimately
degrades the performances of quantum computation. The presence of
static imperfections, although not leading to any decoherence, may
be also detrimental for quantum computers. For instance, a small
inaccuracy in the coupling constants, inducing as a consequence to
errors in quantum  gates, can be tolerated only up to a certain
threshold~\cite{georgeot00}. Moreover, the role of static
imperfections depends on the regime, chaotic or not, of the system
under consideration~\cite{georgeot00}. The stability of a quantum
computation in the presence of static imperfections has been already
analyzed both in terms of
fidelity~\cite{miquel97,georgeot00,benenti01} and
entanglement~\cite{montangero03}.

A strict separation in ``static'' imperfections and ``dynamical''
noise may not be always satisfactory. Dynamical noise may be
considered at the same level as static imperfections, if its
evolution occurs on a scale much larger than the computational time.
In Ref.~\cite{benenti01} it was suggested that the effects of static
imperfections can be more disruptive than noise for quantum
computation. In this Letter, we intend to explore this problem in
more details. The model we consider, in spite of its simplicity,
enables one to grasp the interplay between the different time scales
that appear in the problem. We consider each qubit coupled to a
stochastic variable which changes in time with a fixed frequency.
Below a given threshold (frequency), the errors can be considered as
static, and thus can be corrected by using any of the known methods.
The difference between the chaotic and the other dynamical regimes,
found for static imperfections, holds also in the quasi-static case.
We then generalize our results, by proving a theorem that states
that, under general assumptions, in a perturbed system, unitary
dynamical errors are averaged to zero in probability. Our results
might be relevant in the context of the strategies that have been
proposed during the last few years in order to suppress
decoherence~\cite{preskill}.

\underline{Model} - Following~\cite{georgeot00,benenti01}, we
model a quantum computer as a lattice of interacting spins
(qubits). Due to the unavoidable presence of imperfections, the
spacing between the up and down states (external field) and the
couplings between the qubits (exchange interactions) are both
random and fluctuate in time. We consider $n$ qubits on a
two-dimensional lattice, described by the Hamiltonian
\begin{equation}
    H (t) = \sum_{j=1}^n [\Delta_0 + \delta_j(t)] \sigma_z^{(j)} +
    \sum_{\langle i,j \rangle} J_{ij}(t) \sigma_x^{(i)}\sigma_x^{(j)},
    \label{hamil}
\end{equation}
where the $\sigma^{(i)}_\alpha$'s ($\alpha=x,y,z$) are the Pauli
matrices for qubit $i$ and the second sum runs over nearest-neighbor
pairs. The energy spacing between the up and down states of a qubit
is $\Delta_0+\delta_i(t)$, where the $\delta_i(t)$'s are uniformly
distributed in the interval $[-\delta /2,\delta /2]$ and the
$J_{ij}(t)$'s in the interval $[-J,J]$ (zero means and variances
$\delta^2\sigma^2$ and $4 J^2 \sigma^2$, respectively, with
$\sigma^2=1/12$). We model the dynamical noise by supposing that
both $\delta_i(t)$ and $J_{ij}(t)$ change randomly after a time
interval $\tau$. Within the time interval they are constant.

For $J=\delta=0$ the spectrum of the Hamiltonian is composed of
$n+1$ degenerate levels, with interlevel spacing $2\Delta_0$,
corresponding to the energy required to flip a single qubit. We
study the case $ \delta,J \ll \Delta_0$, in which the degeneracies
are resolved and the spectrum is composed by $n+1$ bands. In this
limit the coupling between different bands is very weak.
We assume free boundary conditions and express all the energy
in units of $\Delta_0$ (we choose $\hbar=1$).

In the following we analyze the behavior of the
fidelity~\cite{peres} and error
\begin{equation}\label{eq:fiddef}
    F(t) \equiv |\langle \Psi |U(t) |\Psi  \rangle|^2 \ , \quad
    E=-\ln F \ .
\end{equation}
starting from an initial state $|\Psi\rangle$ which is an eigenstate
of $\sigma_z^{(j)}$ ($j=1,\ldots,n$), $U(t)$ being the unitary
evolution generated by (\ref{hamil}). We concentrate on the central
band of zero total magnetization which is characterized by the
highest density of states and for which one expects that the effect
of noise is most pronounced.

\begin{figure}
    \centerline{\includegraphics[width=\linewidth]{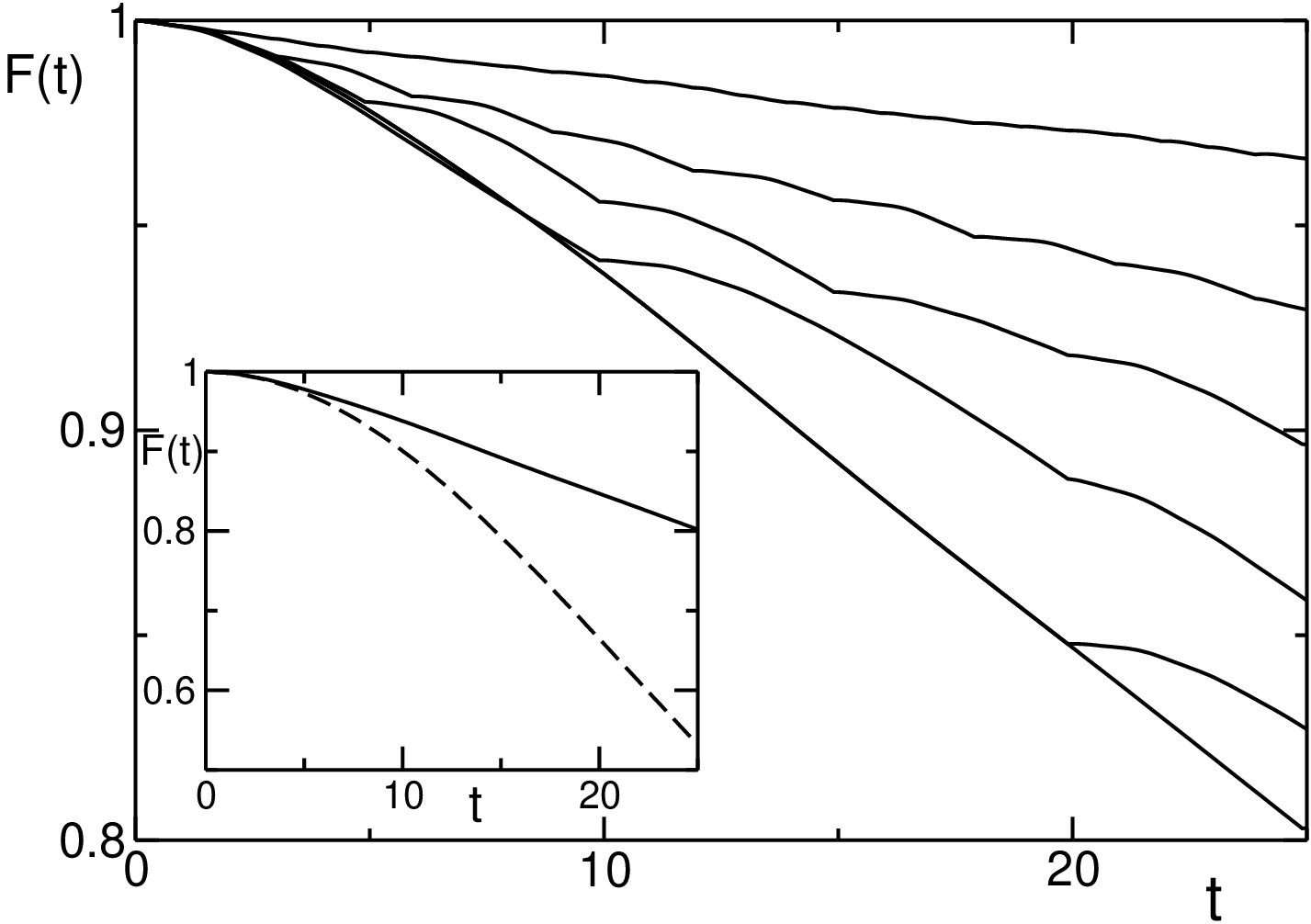}}
    \caption{Fidelity as a function of time for $n=14$ qubits in the
    FGR regime
    ($J=2 \cdot 10^{-2}, \delta=4 \cdot 10^{-1}$) and from top
    to bottom $\tau=
    1,3,5,10,20,25$ (static imperfections). Inset: Fidelity as a
    function of time in the
    ergodic ($J=\delta=2 \cdot 10^{-2}$, dashed line),
    and in the FGR regime (full line).}
    \label{fig1}
\end{figure}

\underline{Results} - The  decay of the fidelity due to static
imperfections  is displayed in the inset of Fig.~\ref{fig1}. The
system (\ref{hamil}) is characterized by two distinct dynamical
regimes depending on the critical value $J_c \sim \delta/n$: the
Fermi Golden Rule (FGR) ($J < J_c$) and the ergodic regime ($J >
J_c$)~\cite{georgeot00, montangero03}. The FGR is characterized by a
Lorentzian local density of states with width $\Gamma^{\rm FGR}$.
The ergodic regime is reached when all the levels inside the band
participate to the dynamics; the local density of states coincides
with the density of states and has a Gaussian shape with variance
$\Gamma^{\rm erg}$.
In both regimes the decay of the fidelity is the Fourier transform
of the local density of states~\cite{flambaum} and follows an
exponential and a Gaussian decay with characteristic decay times
$\Gamma_{\rm FGR} \propto J^2/\delta$ and $\Gamma_{\rm erg} \propto
J^2$ respectively (see Inset Fig.~\ref{fig1})~\cite{georgeot00}.

\begin{figure}
    \centerline{\includegraphics[width=\linewidth]{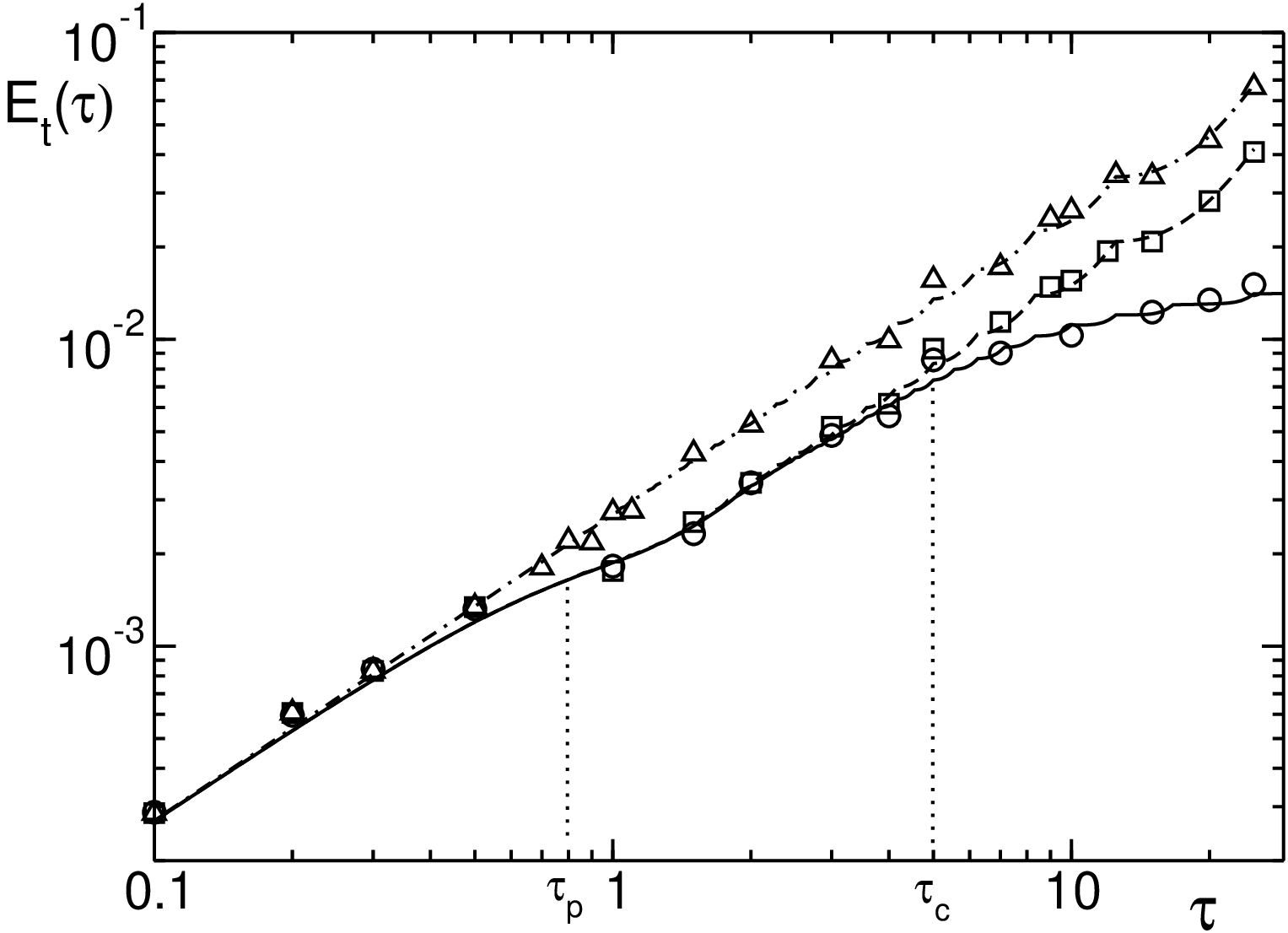}}
    \caption{Error $E$ as a function of $\tau$ for  $t=25, n=10,
    J=5 \times 10^{-3}$, in the ergodic regime $\delta=5 \times
    10^{-3}$, $n_{\uparrow\downarrow}=8$ (squares),
    $n_{\uparrow\downarrow}=13$ (triangles) and in the FGR regime
    $\delta=3 \times 10^{-1}$ (circles).
    The fits are given by Eqs.~(\ref{eq:EN}) and (\ref{eq:erg})-(\ref{eq:fgr}) with
    $\sigma^2=1/12$, $n_c=13$, $\Delta_0=1$  and
    $(n_{\uparrow\downarrow},n_{\uparrow\uparrow})$ equal to
    $(8,5)$ (dashed), $(13,0)$ (dot-dashed). The transition at
    $\tau_c$ is shown only in the former case.
    All the errors scale as $J^2$ (data not shown). }
    \label{fig2}
\end{figure}

In the case of dynamical imperfections, different regimes emerge as
a function of the frequency $1/\tau$. Below a critical timescale
$\tau_c$ the different behavior due to the ergodic and FGR regimes
cannot be resolved anymore. This can be clearly seen in
Figs.~\ref{fig1}-\ref{fig2}. A smoother crossover appears at a
higher frequency $1/\tau_p$ (Fig.~\ref{fig2}) when the noise
frequency become comparable with the single qubit natural frequency
($\sim \Delta _0$). The error $E_t(\tau)$ at (fixed) time $t$ tends
to vanish as $\tau$ decreases.

The explicit calculation of the error to order $J^2$ yields
\begin{equation}
    E_t(\tau)=4 J^2 \sigma^2 \left(N g(\tau)+g(\Delta t)\right) ,
    \label{eq:EN}
\end{equation}
where $t=N \tau + \Delta t$, with $N$ integer, $0\leq \Delta t <
\tau$, and
\begin{equation}
    g(\tau)=2\int_0^\tau ds \int_0^s du\; \mathrm{sinc}^2\! (\delta
    u)\left[ n_{\uparrow\downarrow}+n_{\uparrow\uparrow}
    \cos(4\Delta_0 u)\right] \ ,
    \label{eq:gtau}
\end{equation}
$n_{\uparrow\uparrow}$ ($n_{\uparrow\downarrow}$) being the number
of nearest-neighbor parallel (antiparallel) pairs in the initial
state and $\mathrm{sinc}(x)=\sin x /x$. The integration can be
explicitly performed~\cite{integration} although the resulting
analytic expression is not very transparent. Note that, due to the
convexity of $g(\tau)$, the error $E_t(\tau)\leq 4 J^2\sigma^2 t
g(\tau)/\tau$, being equal when $t/\tau=N$, thus providing a simple
interpolation of (\ref{eq:EN}). Moreover, the function $g(\tau)$ can
be approximated in several important limits. For $\tau\delta\ll1$,
$\mathrm{sinc}^2\! (\delta u)\simeq 1$, whence
\begin{equation}
    g(\tau)\simeq \tau^2\left[
    n_{\uparrow\downarrow}+n_{\uparrow\uparrow} \mathrm{sinc}^2\!
    (2\Delta_0 \tau)\right] \ ,
    \label{eq:erg}
\end{equation}
which yields $g(\tau)\simeq n_c \tau^2$  for $\tau \lesssim
\tau_p=\pi/4\Delta_0$ and $g(\tau)\simeq n_{\uparrow\downarrow}
\tau^2$ (ergodic regime) for $\tau \gtrsim \tau_p$ (see
Fig.~\ref{fig2}), where the total number of links
$n_c=n_{\uparrow\downarrow}+n_{\uparrow\uparrow}$, unlike
$n_{\uparrow\downarrow}$ and $n_{\uparrow\uparrow}$, does not depend
on the initial state $|\Psi \rangle$. On the other hand, when
$\tau\delta\gg 1$, $\mathrm{sinc}^2\! (\delta u)\simeq
\frac{\pi}{\delta} \delta(u)$ (FGR regime) and (\ref{eq:gtau}) reads
\begin{equation}
    g(\tau)\simeq
    n_{\uparrow\downarrow}\frac{\pi}{\delta^2} \left[\delta\tau -\ln(2\delta\tau)-\gamma-1\right] \;
    ,
    \label{eq:fgr}
\end{equation}
where $\gamma\simeq0.577$ is Euler's constant. Substituting these
approximate expressions in Eq.~(\ref{eq:EN}), the error at a fixed
time $t$ for different $\tau$ values scales like
\begin{equation}
    E_t(\tau)\simeq 4 J^2 \sigma^2 t \left\{
    \begin{array}{lll}
    n_c \tau  & \tau < \tau_p & \mbox{(all regimes)}\\
    n_{\uparrow\downarrow} \tau  & \tau_p<\tau < \tau_c
        & \mbox{(all regimes)}\\
    n_{\uparrow\downarrow}\tau &   \tau > \tau_c, \;
        J \simeq \delta & \mbox{(ergodic)} \\
    n_{\uparrow\downarrow}\pi/\delta &   \tau > \tau_c, \; J <
    \delta/n & \mbox{(FGR)}
    \end{array}
    \right.
    \label{scalings}
\end{equation}
In Fig.~\ref{fig2} we show the scaling of $E_t(\tau)$ with $\tau$
for different values of $J$. For the ergodic regime we choose
$J=\delta$, while the FGR is characterized by $J \ll \delta $. 
As $\tau < \tau_c$ the two distinct ergodic and FGR behaviors
of the static case (compared in Fig.~\ref{fig2} only for the
sets with $n_{\uparrow\downarrow}=8$) are not
resolved. Equations (\ref{eq:erg}) and
(\ref{eq:fgr}), plotted in Fig.~\ref{fig2}, are in excellent
agreement with the numerical results. The additional kink at $\tau
\simeq \tau_p=\pi/4\Delta_0$ sets in when single spin dynamics
starts to play a role. We also checked that $\tau_p$ is independent
on $J$ and $\delta$, in agreement with Eq.~(\ref{eq:erg}). The
transition at $\tau= \tau_c$ is striking and occurs when the error
starts deviating from the linear behavior given by
Eq.~(\ref{scalings}). In fact, the crossover between the two regimes
could be defined by equating the third and the fourth line
of~(\ref{scalings}), that is for $\tau=\pi/\delta$, which for
$\delta=0.3$ would give $\tau\simeq10.5$. However, since the
saturation value $E_t(\tau)=4 J^2 \sigma^2 n_{\uparrow\downarrow}
\pi t/\delta$ given by Eq.~(\ref{scalings}) is reached only for
$\delta \tau \gg 1$ and since the transition is sharp, a much more
accurate way to define $\tau_c$ is by looking at the point for which
the deviation from the linear behavior [third line in
Eq.~(\ref{scalings})] becomes apparent. To this purpose we keep the
next-leading correction to Eq.~(\ref{eq:erg}) and  approximate
$\mathrm{sinc}^2\! (\delta u)\simeq 1 - (\delta u)^2 /3$ (for
$\tau\lesssim 1/\delta$) in the integral (\ref{eq:gtau}). For
$\tau\gtrsim\tau_p$ we obtain
\begin{equation}
    g(\tau)\simeq \frac{n_{\uparrow\downarrow}}{\delta^2}
    \left[ (\delta\tau)^2 -
    \frac{(\delta\tau)^4}{18} \right]  \, .
    \label{eq:ergcorr}
\end{equation}
If the plot resolution in Fig.~\ref{fig2} is some fraction $\varepsilon$ of
the total vertical range $4J^2 \sigma^2 n_{\uparrow\downarrow}
t^2$, the error curve starts deviating from the linear behavior
when $(t/\tau) (\delta\tau)^4/(18\delta^2)\simeq \varepsilon t^2$,
i.e.\
\begin{equation}
    \tau_c =\frac{(18\varepsilon
    t)^{\frac{1}{3}}}{\delta^{\frac{2}{3}}},
    \label{tauscal}
\end{equation}
which for $t=25$, $\delta=0.3$ and $\varepsilon=1/40$ yields
$\tau_c=5$, in full agreement with Fig.~\ref{fig2}.
\begin{figure}
    \centerline{\includegraphics[width=\linewidth]{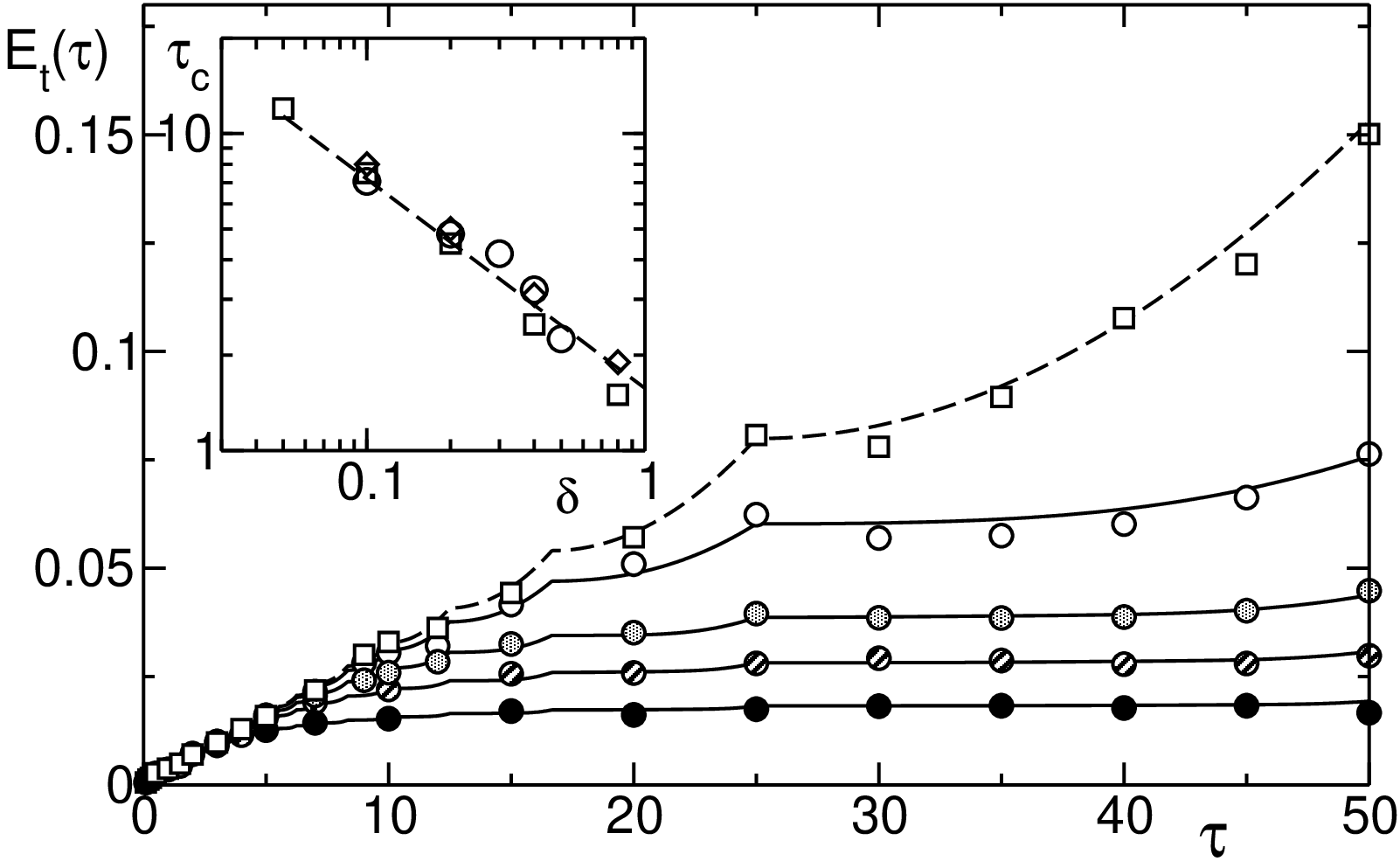}}
    \caption{ Error at time $t=50$, for $n=10, J=5 \cdot 10^{-3}$ and
    different $\delta$ values. The squares represent the ergodic
    regime $\delta=J$. The FGR regime is plotted for $\delta= 1,2,3,5
    \cdot 10^{-1}$ (empty, pointed, dashed, full circles
    respectively). Inset: $\tau_c$ as a function of $\delta$ for
    $n=10,12,14$ (circles, squares and diamonds respectively). The
    dashed line is proportional to
    $\delta^{-2/3}$, in agreement with Eq.~(\ref{tauscal}).
   }
    \label{fig3}
\end{figure}
In Fig.~\ref{fig3} we show the error $E_t(\tau)$ with fixed $J$
and different $\delta$ values. The scaling of the critical
threshold $\tau_c$ is clearly visible. We also checked that
$\tau_c$ does not depend on $J$ (data not shown). The inset of
Fig.~\ref{fig3} shows the dependence of $\tau_c$ as a function of
$\delta$, confirming the prediction (\ref{tauscal}).

\underline{Theorem} - After having presented the overall picture of
dynamical imperfections on the fidelity of computation, we complete
our analysis and set up a general framework to
consider the effect of a time-dependent noise on the evolution
of a quantum system
\begin{equation}
H(t) = H_0 + \xi(t) V ,\label{hamth}
\end{equation}
where $H_0$ is time independent and $H(t)$ varies with a given
characteristic time $\tau$, according to the stochastic process
with independent increments $\xi(t)=\sum_{k=1}^N
\chi_{[k\tau-\tau, k\tau)}(t)\; \xi_k$, where $\chi_A$ is the
characteristic function of the set $A$ and $\{\xi_k\}_k$ are
independent and identically distributed random variables, with
expectations $ E[\xi_k]=0,~ \mathrm{Var}[\xi_k]=
E[\xi_k^2]=\sigma^2 < \infty. $ The time evolution operator over
the total time $t=\tau N$ is given by
\begin{eqnarray}\label{eq:UNT}
U_N(t)= \prod_{k=1}^{t/\tau} \exp[- i (H_0+\xi_k V) \tau] ,
\label{u}
\end{eqnarray}
where a time-ordered product is understood, with earlier times
(lower $k$) at the right. Let us assume, for simplicity, that
$H_0$ and $V$ are bounded operators, so that $U(t)$ is a
norm-continuous one-parameter group of unitaries and all our
subsequent estimates are valid in norm. We are interested in the
existence and form of the limiting time evolution operator
$U_N(t)$ for $N\to\infty$. When expanding the product, one finds
that the term independent of $t$ is $1$, while the term
proportional to $t$ reads
$-i H_0 t -i V t \sum_{k=1}^N \xi_k/N$.
Now, according to the weak law of large numbers \cite{Cheb},
$P\!-\!\lim_{N\to\infty} \sum_{k=1}^N \xi_k/N=E[\xi_k]=0$,
for we assumed $E[\xi_k^2]=\sigma^2<\infty$,
and the limit is taken in probability.
Therefore, for $N\to\infty$
\begin{equation}\label{limt1}
1-i H_0 t -i V t \frac{1}{N}\sum_{k=1}^N
\xi_k\stackrel{P}{\longrightarrow}1-i H_0 t.
\end{equation}
Analogously, by using the weak law of large numbers, one can prove
that all higher powers of $Vt$ vanish in the limit, thus obtaining
\begin{equation}
U(t)\equiv P\!-\!\!\!\lim_{N\to\infty} U_N(t)= \exp(-i H_0 t),
\label{eq:evolprob}
\end{equation}
in the following sense
\begin{equation}\label{evolprobeps}
\lim_{N\to\infty} P\left(\|U_N(t)-\exp(-i H_0 t)\|\ge
\varepsilon\right)=0,
\end{equation}
uniformly in each compact time interval. If the term $\xi(t)V$ is
viewed as exemplifying the effect of (dynamical) error-inducing
disturbances, the above result physically implies that the effects
of the errors are wiped out if their characteristic frequency
$\tau^{-1}$ is sufficiently fast. This defines the purely
dynamical regime.

Another viewpoint can also be adopted, that is somewhat
complementary to the above one. Given a characteristic frequency of
the noise, it is possible to establish an \emph{effective} value of
the strength of the imperfections so that the above result holds
(approximately). In this sense, a natural question is what happens
for large but \textit{finite} $N$. This (physical) question can be
answered by remembering that under the same hypotheses, according to
the central limit theorem, the limiting random variable
$\eta=\lim_{N\to\infty}  \sum_{k=1}^N \xi_k/\sqrt{N}$ exists and is
Gaussian with mean $E[\eta]=0$ and variance $E[\eta^2]=\sigma^2$,
namely it is distributed like
$f(\eta)=(2\pi\sigma^2)^{-\frac{1}{2}}\exp(-\eta^2/2\sigma^2)$.
Thus, by following the same steps that led to (\ref{eq:evolprob}) we
find that for $N\gg 1$
\begin{eqnarray}
    U_N(t)\sim \exp\left[- i \left( H_0 + \eta V/\sqrt{N}
    \right)t\right].
\label{eq:expansiongauss}
\end{eqnarray}
Equation (\ref{eq:expansiongauss}) implies then that for
\emph{fixed} $\tau$, the system ``feels" an effective interaction
strength $\epsilon_{\rm eff}= \sigma \|V\|/\sqrt{N}\propto\sigma
\|V\|\sqrt{\tau}$.

For intermediate values of $N$, Eq.~(\ref{eq:expansiongauss}) is
no longer valid, because it hinges upon the commutativity of $H_0$
and $V$. However, by assuming
that $V\ll H_0$ (e.g.\ in norm), a straightforward expansion shows
that the perturbation $V$ is replaced by
\begin{equation}\label{eq:meanV}
    \bar V(\tau)=\frac{1}{\tau}\int_0^\tau dt\; e^{i H_0 t} V e^{-i
    H_0 t} \ ,
\end{equation}
so that, for $\tau \|H_0\|\gtrsim 2\pi$, the effective
perturbation becomes
\begin{equation}\label{eq:VZ}
    \bar V(\tau) \to V_Z=\sum_k P_k V P_k ,
\end{equation}
where $P_k$ are the eigenprojections of $H_0$ ($H_0=\sum \lambda_k
P_k$). This phenomenon is reminiscent of the quantum Zeno
subspaces \cite{theorem}.

 The generalization of the above results
to a Hamiltonian with a family of independent stochastic processes
with zero mean and finite variances is straightforward.
This is the case of the Hamiltonian (\ref{hamil}), which reads
\begin{equation}\label{eq:vectorham}
    H(t)=H_0+ \delta\; \bm{\xi_0}(t)\cdot \bm{V_0}+2 J\;
    \bm{\xi}(t)\cdot \bm{V} ,
\end{equation}
where $H_0=\sum_j \Delta_0 \sigma_z^{(j)}$,
$(\bm{V}_0)_j=\sigma_z^{(j)}$,
$(\bm{V})_{ij}=\sigma_x^{(i)}\sigma_x^{(j)}$, and
$\xi_{0j}=(\bm{\xi}_0)_j$ and $\xi_{ij}=(\bm{\xi})_{ij}$
$(i,j=1,\cdots,n)$ are independent random variables uniformly
distributed in the interval $[-1/2, 1/2]$.

We can then reinterpret our previous results in the light of the
above theorem, by applying the well-known static results
\cite{flambaum} to the (static) evolution with renormalized
couplings (\ref{eq:expansiongauss}) (with $\eta V \rightarrow
\bm\eta\cdot\bm V$). Thus, independently of the interaction
strength and the correspondent dynamical regime, there is a
quadratic decay law for sufficiently large $N=t/\tau$ (or small
$\tau$),
\begin{equation}
    \label{quadratic}
    E_t(\tau) \sim \frac{1}{N}\frac{t^2}{\tau_Z^2}=
    \frac{t}{\tau_Z^2}\tau \quad (\tau < \tau_p) ,
\end{equation}
where $\tau_Z^{-2}=4 J^2 \langle\Psi|(\bm{\eta}\cdot
\bm{V})^2|\Psi\rangle= 4 J^2 n_c \sigma^2$ and $\tau_p \simeq
\Delta_0^{-1}$ [the $H_0$-timescale, see (\ref{eq:meanV})]. On the
other hand, for smaller $N$, i.e.\ $\tau>\tau_p$, the effective
interaction (\ref{eq:VZ}) is given by $(\bm{V}_Z)_{ij}
=\sigma_+^{(i)}\sigma_-^{(j)}+\sigma_-^{(i)}\sigma_+^{(j)}$,
whence
\begin{equation}
    E_t(\tau)\sim\frac{1}{N} \Gamma_{\mathrm{erg}}t^2=
    \Gamma_{\mathrm{erg}}t\tau  \quad (\tau
    > \tau_p),
    \label{eq:FNerg}
\end{equation}
where $\Gamma_{\mathrm{erg}}=4 J^2 \langle\Psi|(\bm{\eta}\cdot
\bm{V}_Z)^2|\Psi\rangle= 4 J^2 n_{\uparrow\downarrow} \sigma^2$.
Therefore, we recover the linear growth of the error (with the
correct coefficients), that describes both regimes up to $\tau_c$
in Eq.~(\ref{scalings}).

\underline{Conclusions} - We studied the effects of dynamical
imperfections on a general model of a quantum computer and
identified several dynamical regimes, depending on the frequency of
the external noise as compared with the coupling constants of the
quantum computer. Above a threshold frequency, imperfections can be
treated as static imperfections, although with renormalized
parameters. Below this threshold the different dynamical regimes
induced by the presence of imperfections are not resolved. These
results give a better comprehension of the general problem of noise
in quantum computers and might suggest new strategies to develop
general error correcting techniques.

This work was supported by the European Community under contracts IST-SQUIBIT,
IST-SQUBIT2 and RTN-Nanoscale Dynamics.

\end{document}